\magnification =\magstep1     % determines size of characters
\baselineskip = 16pt          % determines spacing between two lines
\raggedbottom
\vsize = 9.0 truein
\hsize = 6.5 truein
\voffset = 0.0 truein
\hoffset = 0.0 truein
\nopagenumbers
\def\hangpara{\par\hangindent 25pt\noindent}

\def\lya{{Ly$\alpha$\ } }
\def\cm2{cm$^{-2}$}
\parindent = 25 truept
\pageno = 1

\centerline{}
\vskip 0.5 truein
\centerline{\bf THE LYMAN-ALPHA FOREST AT $z\sim$4: } 
\centerline{\bf KECK HIRES OBSERVATIONS OF Q 0000$-$26} 
\bigskip\bigskip
\centerline{Limin Lu$^{1,2}$, Wallace L. W. Sargent$^1$, 
   Donna S. Womble$^{1,2}$, \& Masahide Takada-Hidai$^{1,3}$}
\vskip 2.0 truein
\centerline{(Accepted by the Astrophysical Journal)}
\vskip 2.0 truein

\noindent$^1$ Palomar Observatory, California Institute of Technology, 
105-24, Pasadena, CA 91125

\noindent$^2$ Hubble Fellow

\noindent$^3$ On sabbatical leave from Research Institute of Civilization, 
  Tokai University, Hiratsuka, Japan
\eject
\headline={\hss -- \folio\ -- \hss}
\centerline{\bf ABSTRACT}
\bigskip

    We derive H I column density and Doppler width distributions for 
a sample of
\lya clouds with $3.4<z<4.0$, using a high resolution spectrum of the
quasar Q 0000$-$26 ($z_{em}=4.127$) obtained with the Keck telescope.
Simulated \lya forest spectra with matching characteristics are 
analyzed  similarly in order to gauge the effects of line 
blending/blanketing and noise in the data.  The H I column density 
distribution, after corrections for incompleteness resulting from 
line blanketing, is well described by a single power law function
with index $\beta=-1.55\pm0.05$ over the column density range
of $12.6<{\rm log}N<16.0$. A steepening in
the column density distribution at log $N$(H I)$>14.5$ may be present.
The Doppler width distribution of the clouds
 is consistent with a Gaussian function
with a mean of 23 km s$^{-1}$ and a dispersion of 8 km s$^{-1}$, but 
with a cutoff at 15 km s$^{-1}$, {\it ie},
no clouds with $b<15$ km s$^{-1}$ are required to describe the data.
While the H I column density distribution found here is consistent with
that derived from similar quality data at lower redshifts, 
both the mean Doppler width and the cutoff value
are smaller than those found at lower redshift.
There is a hint for clustering in the clouds' line of sight distribution
in the velocity interval $100<\Delta v<160$ km s$^{-1}$, 
but the evidence is only
marginal.  Analyses of the proximity effect indicate a  value of 
$J_{\nu}^{LL}\sim2\times10^{-22}$ erg s$^{-1}$ cm$^{-2}$ Hz$^{-1}$ 
sr$^{-1}$ for the mean intensity of the metagalactic
UV ionizing background at $z\sim 4.1$, which is consistent with that
expected from high-redshift quasars.

\bigskip
\noindent{\it Subject heading:} diffuse radiation - intergalactic medium -
quasars: absorption lines - quasars: individual (Q 0000$-$26)

\eject
\centerline{\bf 1 INTRODUCTION}
\bigskip

    Studies of \lya absorption lines in spectra of quasars have seen
rapid progress recently. On the one hand,  the Hubble Space
Telescope now allows for investigations of \lya clouds over the important
redshift region $z<1.6$ (eg. Bahcall et al. 1993; 1996), while the 
advent of large aperture
telescopes has greatly facilitated detailed observational studies of
\lya forest absorption at $z>1.6$ at resolutions and S/N considerably higher
than previously possible (eg. Hu et al. 1995; Tytler et al. 1995). 
On the other hand, advances in computing technology now allow for
large-scale numerical simulations of structure formation in the early
universe in various cosmological models.
In particular, several recent papers (Cen et al. 1994;
Petitjean, Mucket, \& Kates 1995; Zhang, Anninos, \& Norman 1995; 
Hernquist et al. 
1996; Miralda-Escude et al. 1996) have  explored the implications
of these cosmological models for the nature of the \lya absorption clouds.
The results suggest that most \lya clouds at high redshift ($2<z<4$)
are likely associated with density
enhancement in relatively low density regions of the universe in between
collapsed structures.  All these model calculations, 
with their different choices of cosmological parameters,
different treatment of the physics involved, and different 
computational techniques,
appear to reproduce the basic properties of \lya clouds as 
inferred from observations. Thus
more careful scrutiny of the models and more accurate knowledge of
the \lya cloud properties from observations are necessary in order to
decide which cosmological model is correct.

   In this paper, we present observations and analyses of the \lya
forest spectrum of the $z_{em}=4.1$ quasar Q 0000$-$26 obtained with the
Keck 10m telescope. The main aim is to derive accurate distributions of
cloud parameters (redshift, H I column density, Doppler width), which can
then be compared to similar distributions estimated for lower
redshift clouds in order to study the evolution of the cloud properties.
The study will also provide reliable observational reference against
which theoretical models can be tested. Simulated \lya 
forest spectra with matching characteristics to the observation
are also analyzed in the same way in order to understand
any biases that might exist in the derived parameter distributions
resulting from line blending/blanketing, noise in the data, and the 
analysis technique itself.

    The organization of the paper is as follows.
After describing the observations and data reduction in \S2, we
discuss the profile fitting technique used to estimate $z$, $N$(H I), and
Doppler $b$ for the \lya clouds (\S3) and the numerical simulations
(\S4). The main observational results are presented in \S5, including
the distributions of column density and Doppler width, the clustering
property of the clouds along the line of sight, and a proximity effect
analysis. The results obtained here are compared to results obtained
in similar studies for lower redshift clouds and to recent theoretical
models in \S6.  A summary of the main results is given in \S7.

\bigskip
\centerline{\bf 2 OBSERVATIONS AND DATA REDUCTION}
\bigskip
    We observed Q 0000$-$26 on 13 November 1993 with the High 
Resolution Echelle Spectrometer (HIRES) of the Keck 10m telescope 
(Vogt 1992).  A 0.86" slit width was used  to achieve a resolution
of FWHM=6.6 km s$^{-1}$ with roughly 3 pixels per resolution element.
Four exposures of 3000 sec each were obtained. Complete spectral coverage
in the range of 5160-7580 \AA\ was obtained with two partially overlapping
setups. Data reductions were done in the usual manner using a software
written by T. A. Barlow.  After the echelle orders were extracted
and wavelength- and flux-calibrated, they were
resampled to a uniform wavelength scale (but keeping roughly
the same number of pixels in a resolution element) and added together
according to their S/N. The resulting spectrum has a typical S/N 
(per resolution element) in the \lya forest region 
of 25-35:1, corresponding to a
$4\sigma$ detection limit of $N$(H I)$\sim 2.5\times10^{12}$ \cm2 
for an isolated \lya line with a Doppler width $b=20$ km s$^{-1}$ 
($b$ is related to 
the velocity dispersion $\sigma$ through the relation $b=\sqrt{2}\sigma$).
The S/N longward of \lya emission decreases from $\sim 80:1$ near the
peak of the \lya emission to 20-30:1 in the last echelle order covered.

    The continuum level longward of the \lya emission was established
by fitting cubic splines to regions deemed  free of absorption features
using the IRAF task {\it continuum}. Such a procedure did not work very well
in the \lya forest region as most parts of the \lya forest
show no obvious continuum due to heavy absorption. We estimated the
continuum level in the \lya forest by picking out small ``peaky'' regions 
(typically a few \AA\ wide) in the \lya forest
 that are free of obvious absorption features
and connecting them with straight lines. Some hand-editing of the 
continuum near the \lya emission and the damped \lya absorption 
at $z_{abs}=3.39$ (see Figure 1) were 
necessary. The resulting continuum appears
to describe the data satisfactorily. Figure 1 shows the \lya forest portion
of the HIRES spectrum along with the estimated continuum level.
The continuum-normalized spectrum is
shown in Figure 2 in finer detail, along with the 1$\sigma$ error spectrum.
Note that the damped Ly$\alpha$ absorption at $z_{abs}=3.39$ has already 
been removed from Figure 2.
This damped Ly$\alpha$ absorption system will be analyzed elsewhere.

    Absorption lines longward of the Ly$\alpha$ emission are selected
using an automated software as described in Tripp, Lu, \& Savage (1996), 
which yields a central wavelength, a measured equivalent width, and a
1$\sigma$ error of the measured equivalent width for each
absorption feature. In general, only features over the 4$\sigma$ 
significance level
are retained in the line list (Table 1). In a few cases,
absorption features with significance level between 3 and 4$\sigma$ that
apparently correspond to lines in identified metal absorption systems
are also retained and noted in the line list.
    We have attempted to identify all lines longward of the \lya emission.
These identifications are also given in Table 1. There are a
number of weak ($<6\sigma$) features which have no obvious identifications. 
We suspect most of these are C IV$\lambda$1548 absorption 
lines whose corresponding weaker doublet members
are  below the $3\sigma$ significance level. Lines occurring
in the \lya forest that are associated with identified metal systems
are also indicated in the line list. We did not attempt to identify new 
metal systems (eg, C IV doublets) based solely on lines in the \lya forest 
as the probability for chance coincidence is expected to be very high. Based
on the statistics of weak C IV systems: $dN/dz=7.1$ for $w_r\geq 0.03$ \AA
(Tripp et al. 1996), we expect roughly 4 C IV doublets with 
$w_r\geq0.03$ \AA\ in the \lya forest of Q 0000$-$26. Thus the level of 
contamination should be very small given that there are several hundred
\lya lines in the observed \lya forest.

\bigskip
\centerline{\bf 3 VOIGT PROFILE FITTING}
\bigskip
     At the resolution employed (FWHM=6.6 km s$^{-1}$), 
all \lya lines are resolved.
To derive the redshift, H I column density, and Doppler width
for each individual \lya cloud, we fit Voigt profiles to all \lya lines
between 5380 \AA\ and the \lya emission. The lower cutoff of the fitting
region is imposed by the damped \lya absorption at $z_{abs}=3.39$.
Although
we have a small coverage ($\sim 100$ \AA) of the region below the Ly$\beta$ 
emission line, where a number of Ly$\beta$ absorption lines can be
identified, we have decided not to use the Ly$\beta$ lines to constrain
the profile fitting in order to achieve uniform treatment of the entire
line sample. In general, the Ly$\beta$ lines are contaminated
by \lya forest lines at lower redshifts and their usefulness is limited.

     The software we use to fit Voigt profiles is 
VPFIT  developed by R. F. Carswell and collaborators and kindly
made available to us. It is a
$\chi^2$-minimization program which estimates simultaneously the 
redshift ($z$),
column density ($N$ or $N$(H I)), Doppler width ($b$) and associated errors
for each component (cloud) in the fitting region. 
At the high redshifts we are dealing with,
most \lya lines are more or less blended with neighboring lines.
Our fitting strategy is to divide the forest region into many sections
where the two ends of each section recover to the
continuum level, and to fit each section separately. The size of the
sections ranges from a few angstroms in relatively uncrowded regions
up to 50 \AA\ in heavily blended regions, while the number of lines in
the sections varies from a few to several tens. Clearly
one needs a convenient, meaningful, and objective way to decide how 
many components to insert in fitting a section. We start
with the minimum number of lines (components) 
recognizable to the eye, and add more lines as needed until a reduced
$\chi^2$ of $\leq 1.1$ per degree of freedom is achieved. 
Occasionally this ``rule'' has to be relaxed in order 
to accommodate spectral regions where cosmic ray events or defects on
the detector not removable by the flatfielding procedure have
apparently corrupted the spectrum (eg, regions around 5595, 5867, 
and 5955 \AA).
We also examine the residuals of the fits and find that the above
stopping criterion almost always yields visually appealing results
(see figure 2).
    
    Identified metal lines in the \lya forest are also fitted using 
similar procedures in order to recover any Ly$\alpha$ lines blended with 
the metal absorption lines. Lines from the same ion species 
(eg, Si IV$\lambda\lambda$ 1393, 1402) in a given system
are fitted simultaneously with their parameters tied together.

The final list of \lya and metal lines resulting from the Voigt
profile fitting process is given in Table 1, where we give a line
number for identification, central wavelength of the absorption,
identification, redshift, column density, and Doppler width, along
with the $1\sigma$ errors where appropriate. 
The VPFIT results are also shown in Figure 2 superimposed on the
observed spectrum.
Only the \lya lines will be discussed in the remainder of this paper.
The metal systems will be discussed in a future paper when combined
with spectra of other quasars.

\bigskip
\centerline{\bf 4 NUMERICAL SIMULATIONS}
\bigskip

     Most weak \lya lines are expected to be lost in the observed
spectrum due to heavy
line blanketing.  In order to derive unbiased
distributions of column density and Doppler width, we 
generate simulated \lya forest spectra from computer codes
and fit the simulated spectra in the same way as we do for the
observed spectrum. This procedure should help us to understand
and, in some cases, to correct for effects that are caused by
line blending/blanketing and/or the Voigt profile fitting process itself.

    The simulated spectra have the same spectral
coverage, resolution, sampling rate (pixel size), and S/N as the
observed one. \lya lines are drawn randomly from given distributions
in $z$, $N$, and $b$. The assumed distribution is:
$$f(z,N,b)=A(1+z)^{\gamma}N^{-\beta}
   e^{-({ {\langle b\rangle-b} \over \sqrt{2}\sigma })^2} \eqno(1) $$
for $N_{min}\leq N\leq N_{max}$ and $b_{cut}\leq b\leq b_{max}$. 
This functional form has been found to give reasonable descriptions
of the observed properties of Ly$\alpha$ clouds (cf. Carswell 1988).
In order for the simulations
to be as realistic as possible, all metal lines that occur in the 
observed \lya forest of Q 0000$-$26 (including those which are 
expected to occur in the \lya forest
but are not recognizable in the observed spectrum because of blanketing)
are inserted back into the simulated spectra using the parameters
derived from VPFIT (section 3 above). A
proximity effect (cf. Bajtlik, Duncan, \& Ostriker 1988; see also \S5.5)
is also put in the simulation using the parameters
given in \S5.5. The normalization constant in equation (1)
is adjusted such that the resulting simulated spectra have
the same mean $D_{A}$ value (=0.511 between [5400, 6100] \AA) as the 
observed spectrum (see \S5.5 for an explanation of $D_A$).

   We fix the value of $\gamma$ at 2.75 (Lu, Wolfe, \& Turnshek 1991).
For the small redshift range considered here, the exact value of $\gamma$
is not important. We also choose $b_{max}=100$ km s$^{-1}$ based on the
results of previous studies (cf. Carswell 1988), and set $N_{max}=10^{18}$
cm$^{-2}$ in order to avoid damped Ly$\alpha$ absorption lines.
The remaining parameters in equation (1) ($\beta$, $N_{min}$,
$\langle b\rangle$, $\sigma$, and $b_{cut}$) are then varied in order
to explore the parameter space. The resulting simulated spectra are subjected
to several simple tests to select out the ``best candidates'' for profile 
fitting and for further comparison with the observed spectrum. 
In the first test, we compare the intensity distribution of the pixels
between the simulated spectra and the observed spectrum
in the spectral region [5400, 6100] \AA. In the second test, 
we compare the power spectrum of the simulated spectra and
the observed spectrum in the same spectral region.
The first test appears to
be reasonably sensitive to the column density distribution, while the
second test appears to be more sensitive to the Doppler width distribution.
After this preliminary screening, simulations which compare favorably
with the observed spectrum are fitted with Voigt profiles using the
same procedure as described in \S3, and the results are compared
with those from fitting the observed spectrum. The parameters that
match the observation the best are given in Table 2.
In the remaining discussion, unless noted otherwise, the results of 
profile fitting one arbitrary simulated spectrum using the 
parameters given in Table 2 will be compared to the results 
of profile fitting the observed Q 0000$-$26 spectrum.

\bigskip
\centerline{\bf 5 RESULTS}
\bigskip

     \lya lines near quasar emission redshifts are affected
by the proximity effect (cf. Bajtlik et al. 1988),  and they should not be 
used in deriving unbiased distributions of cloud parameters. Thus
we define {\it Sample 1} as all \lya clouds within the
redshift range $3.425543\leq z \leq 3.976680$, corresponding
to the wavelength range $5380\leq \lambda\leq 6050$. The upper
cutoff at 6050 \AA\ is chosen because this is where the
UV ionizing intensity from the quasar is expected to roughly equal 
that of the UV background (section 5.5). There are 336 Ly$\alpha$
lines in Sample 1 for the real data, and 373 lines for the simulation.
The mean redshift of the sample is $\langle z\rangle=3.7$.

     For some purposes, it is useful to have a sample where
the measurements of cloud parameters from VPFIT are relatively well
determined so that scatters due to noise/blending would be small.
Thus we define {\it Sample 2} as a subsample of Sample 1 with
measurement uncertainties $\sigma_{{\rm log}N}\leq 0.1$ and
$\sigma_{b}\leq 5$ km s$^{-1}$ from profile fitting. Sample 2 contains
227 lines for the real data and 235 lines for the simulation.
Note that none of the \lya lines with $N>10^{16}$ from the profile fitting
is retained in Sample 2. These lines are strongly saturated and their 
measured parameters from VPFIT are extremely uncertain.

     It is worth pointing out that \lya lines associated with identified
metal systems are included in the above samples if they satisfy the
selection criteria. While traditionally the metal systems and
\lya clouds which do not show obvious metal absorption
have been treated differently since they may belong to different 
parent populations (cf. Sargent et al. 1980), recent studies suggest
that they may in fact be related. For example, Lu (1991)
showed using a composite spectrum that many of the 
previously-thought-metal-free \lya clouds actually contain metals. This  is
corroborated by recent high quality Keck observations
(Cowie et al. 1995; Tytler et al. 1995; Sargent et al. 1996).
In particular, these authors concluded 
that roughly half of the \lya clouds with
$3\times 10^{14}<N$(H I)$<10^{15}$ \cm2  show C~IV absorption,
while essentially all \lya clouds with $N$(H I)$>10^{15}$ \cm2 show C IV 
absorption.  The limiting $N$(H I) above which Ly$\alpha$ clouds are 
found to show detectable metal absorption is still limited
by the sensitivity of the data. Hence, in principle, all \lya clouds 
{\it could}
contain metals. Additionally, some authors (cf. Cristiani et al. 1995)
have found evidence
for clustering in the ``traditional'' \lya forest clouds, further blurring
the differences between metal systems and the ``traditional'' \lya forest
clouds. We will therefore not make the distinction between metal systems
and \lya forest clouds {\it with respect to the
\lya absorption} for the purpose of this work. 
It is important to note that this choice makes
little difference to the statistical results
as the number of identified metal systems is very small
compared to the number of forest clouds.

\bigskip
\centerline{\it 5.1 Column Density - Doppler Width Relation}
\bigskip

       In Figure 3 we plot $b$ vs log$N$ for \lya clouds
in Sample 1 for Q0000$-$26 and for an arbitrary
 simulation using the parameters given in Table 2 for which we have 
performed profile fitting. Figure 4 shows the equivalent for Sample 2,
but with error bars.

       There are a number of interesting features in the $N$-$b$ 
distribution that are worth mentioning. First,
we note the remarkable similarity between the simulation and
the actual data.
Secondly, we note the voids in the upper-left corner of the $N-b$
distribution for both the real data and the simulation.
Since the input cloud distribution to the simulation contains 
many clouds in this region, the aforementioned void 
apparently results from 
line blending and the limited S/N of the data: these
lines are wide and shallow and thus are easier to miss. We also note
another artifact in the $N$-$b$ distribution: the presence of
very narrow ($b<15$ km s$^{-1}$) lines, especially at low column densities.
These narrow clouds are not  present in the input distribution to the 
simulation, and they are clearly a manifestation of line blending and
noise in the data.  Similar conclusions have been reached by Rauch et al. 
(1992, 1993). The number of these vary narrow lines drops dramatically
in Sample 2 (Figure 4), suggesting that higher S/N should reduce this
bias.   Even though at face value Figures 3 \& 4 may  suggest
significant correlations between $N$ and $b$, the simulation results
indicate that such apparent correlations
are probably artifacts due to line blending and insufficient S/N.
Thus the real cloud distribution is  consistent with no intrinsic correlation
between $N$ and $b$. On the other hand, if indeed there is an intrinsic lack
of Ly$\alpha$ clouds with low-$N$ and high-$b$ in nature, such an
effect will be difficult to recognize for the reasons described above.

    We also note the relative lack of  clouds with $b<15$ km s$^{-1}$
in the real data, which is most obvious in Figure 4 
where the scatter due to noise and blending is smaller. 
The agreement in the $b$ cutoff between the
real data and the simulation, whose input cloud distribution has
a cutoff at $b_{cut}=15$ km s$^{-1}$, suggests that
the observed cloud distribution is consistent with having no clouds
with $b<15$ km s$^{-1}$. The few clouds seen with $b<15$ in the real data
are probably either due to noise/blending or are unidentified metal lines
(most likely single members of the C IV doublets with the other
member blended with other lines).  Since we can successfully recover
narrow metal lines put in the simulations (see also Table 1), {\it 
the lack of very narrow Ly$\alpha$
lines in the Q 0000$-$26 spectrum is not an artifact of the profile
fitting process}.

    To further test the reality and accuracy of the cutoff
$b$ value, figure 5 shows the $N$-$b$ relation for another simulation
where the input cloud distribution has a $b_{cut}=18$ km s$^{-1}$.
One can see that the recovered cloud distribution from VPFIT shows much 
better agreement with a cutoff $b$ value of 18 km s$^{-1}$ than with a 
cutoff value of 15 km s$^{-1}$.  Hence we believe the cutoff $b$ value 
of 15 km s$^{-1}$ estimated for the real data is fairly robust.

    There is some evidence that the $b$ cutoff increases 
sightly with H I column density, as suggested in Hu et al. (1995). 
Interestingly,
such a dependence of $b_{cut}$ on column density appears to have
a possible physical origin as demonstrated by the CDM simulations of
Zhang et al. (1995, see their figure 4). This dependence
arises because higher column density clouds appear to be associated with 
denser regions where the gas is hotter
due to the increasing shock velocities (Miralda-Escude et al. 1996).  
A similar dependence of
$b_{cut}$ on $N$(H I) is visible in our data for Q 0000$-$26 
(Figures 3 \& 4, upper panels).
However, we see the same dependence in 
some of our simulations (cf. Figure 5), 
which suggests that this dependence
of $b_{cut}$ with column density can also be created by the
profile analysis procedure itself. Thus we are not confident of the 
reality of this dependence.

\bigskip
\centerline{\it 5.2  Column Density Distribution}
\bigskip

In the upper panel of Figure 6 we show the column density 
distribution from Sample 1.
The solid histogram is for the observation, and the dotted histogram
is for the simulation. The dashed straight line illustrates the input 
distribution to the simulation, which has an index of $\beta=1.53$.
We first note the good agreement between the
distributions from the simulation and from the observation, which is taken
as an 
indication that the simulation parameters must be substantially correct.
We also note the progressively larger discrepancy toward lower
column density between the input cloud distribution and that
recovered from VPFIT to the simulated spectrum. This is apparently
due to the line blanketing effect since the simulated spectrum has
enough S/N to detect most of these ``missing'' lines if they are isolated.
 We have used the simulation to
correct for the incompleteness of the observed distribution, and the
resulting distribution is shown in the lower panel of Figure 6. 
However, the corrections are only made for the
data points at log$N<13.5$ because the distribution
at log$N>$13.5 appears to be largely unaffected by the line blanketing 
effect (in the statistical sense),
and because the bins at higher column densities 
contain fewer lines and the correction
factors are more uncertain.  A $\chi^2$ fit to the corrected $N$(H I) 
distribution yields $\beta=1.55\pm0.05$ for $12.6\leq{\rm log}N\leq16.0$. 
The lowest column density bin was not used in the fit because the correction
factor is very large ($\sim 25$) and relatively uncertain
(this bin is near the detection limit for the quality of our data).
We also tried different
correction regions and fitting regions. These results are summarized
in Table 3. Evidently all fits given in Table 3 are acceptable.

We note that Hu et al. (1995) obtained $\beta=-1.46$ with a 95\% confidence
range of ($-1.37$,$-1.51$) in the column density range 
$12.3\leq{\rm lg}N\leq 14.5$ based on the analysis of similar Keck data, but
at a mean redshift of 2.8.  We obtain $\beta=1.46\pm0.06$ for the same
column density range, in excellent agreement with the Hu et al. 
determination. 

It has been suggested that there may be a steepening in the column 
density distribution of \lya clouds at log$N>14$ 
(Carswell et al. 1987; Giallongo et al. 1996). Similarly,
both Petitjean et al. (1993) and Hu et al. (1995) demonstrated that 
there is a deficit of clouds in the column density range
$14.5<$log$N<17$ compared to a power-law extrapolation of the
distribution from lower column densities. The physical cause of this
deficit is not exactly clear.  We see a hint of such a
deficit in our data (Figure 6), although the deficit is
not very significant since a single power law function yields an 
acceptable fit to the distribution over the entire column density 
range $12.6<$log$N<16.0$.

\bigskip
\centerline{\it 5.3  Doppler Width Distribution}
\bigskip

      In Figure 7 we show the distribution of Doppler parameters for lines
in Sample 1 and Sample 2 for both the observation and the simulation.
We also show with the smooth dotted curves
the input $b$ distribution to the simulation. It is seen that
the observation and the simulation yield very similar Doppler width
distributions. A Kolmogorov-Smirnov test indicates that the observed
$b$ distribution is consistent with that from the simulation: the
probabilities for the two distributions to be drawn from the same
parent population are 16\%  and 49\% for Sample 1 and Sample 2, respectively.

     As discussed in \S5.1, the combination of noise and blending
has created several artifacts. The first of which is the ``creation'' of
clouds with very small $b$ values ($<15$ km s$^{-1}$) at low column 
densities. This effect can be remedied with higher S/N observations.
The second is the ``creation'' of excessive clouds with 
large $b$ values over the input distribution (figure 7), 
some  as large as 100-200 km s$^{-1}$ (see figure 3).
A careful examination of the clouds with estimated
$b>100$ km s$^{-1}$
indicates that they are always associated with region of the spectrum 
where blending from a multitude of lines have pulled the entire spectrum 
below the continuum. Thus these broad features
are chiefly caused by heavily blended forest lines. 
Imperfect placement of the continuum level in the observed spectrum
may also play a role, although its effect is difficult to assess.
In any case, the fact that the Doppler width distributions from 
the real data and from the simulation agree well suggests
that the input Doppler distribution given in equation (1) and
Table 2 is a reasonably good description of the real distribution.

     Tytler et al. (1995) suggested that there are several types of
Ly$\alpha$ lines, including a type of vary narrow lines ($b\sim 3-15$
km s$^{-1}$) and a type of very shallow, broad lines
($b\sim60-200$ km s$^{-1}$). The data used in the Tytler et al. study
are of similar resolution as ours but with $\sim3$ times better S/N.
The existence of such narrow lines
would imply very cool clouds with temperature significantly below
the canonical value of a few$\times 10^{4}$ K. The existence of the very 
broad lines might suggest hot, collisionally ionized gas, which could
contain lots of baryon.  However, our simulation 
results indicate that  such very narrow and very broad
features could be introduced artificially by noise and blending. 
Although we suspect that at least some of the very narrow and very 
broad features found by Tytler et al. are artifacts caused by noise
and blending, an accurate assessment of the their reality will
require a careful analysis of simulated spectra like the kind performed
here but with S/N appropriate for the Tytler et al. spectra.

\bigskip
\centerline{\it 5.4  Clustering Properties}
\bigskip

   The most common way to investigate the clustering properties 
of quasar absorption line systems along the line of sight 
is to construct the two-point
correlation function (Sargent et al. 1980), which is the distribution
of pair-wise line separations in velocity or space.
In Figure 8 we show the two-point correction functions
(solid histogram) of the \lya clouds in Sample 1 against velocity
separation: $\Delta v={(z_2-z_1)c \over 1+<z>}$, where $c$ is the 
speed of light. The upper panel is for the real data and lower panel
for the simulation. The waving continuous curves in each case indicate the
$\pm1\sigma$ standard deviation expected from randomly distributed
(ie, unclustered) clouds as determined from Monte Carlo simulations. 
The lack of any significant clustering signal in the simulation is 
consistent with the input cloud distribution, which is random
other than a general increase of line density with $N(z)\propto(1+z)^{2.75}$. 
The simulation results also
indicate that at velocity separations $\Delta v<50$ km s$^{-1}$
or so, any information on clustering is lost owing to the intrinsic
width of the lines and possibly blending, too. 
There is a weak clustering signal at $100<\Delta v<160$ km s$^{-1}$
in the observed cloud distribution for $13.0<$log$N<15.0$.
Monte Carlo simulations indicate
that the probability of finding 3 consecutive 
velocity bins with deviations at least as large as those observed 
at $100<\Delta v<160$ km s$^{-1}$ in the real data for randomly
distributed clouds is 3.7\% (the
probability doubles if negative deviations or anti-clustering are
also considered).  Thus the statistical significance of this
clustering signal is only marginal. 
The significance of this clustering signal drops for
any other choices of column density range.

\bigskip
\centerline{\it 5.5  Proximity Effect}
\bigskip

     The proximity effect, which is the deficit of \lya clouds near
a quasar emission redshift, can be used to estimate the mean intensity
of the UV ionizing background at that redshift assuming the effect
is caused by the enhanced UV radiation from the nearby quasar (Bajtlik
et al. 1988). This is generally done by comparing the number density 
of \lya lines above a certain completeness limit (in equivalent width 
or column density) near the quasar emission line with that expected from 
the general distribution of the \lya clouds determined from regions
far away from the Ly$\alpha$ emission.
Such an approach, however, is not very practical here because the 
completeness limit in $N$(H I) not only depends on the S/N of the data, but
also on the degree of blending (or equivalently the mean density of lines);
both are very different near the \lya emission.
Figure 6 suggests that if we are to use a constant column density cutoff 
to ensure completeness of the sample, then the cutoff has to be
at least as large as log$N$=13.5, which would severely limit the number of 
lines in the sample. For this reason, we adopt the following approach. 

     We divide the observed spectrum into 100 \AA\ bins starting from
the emission redshift $z_{em}$, and calculate
the mean $D_A$ value of each bin, where $D_A$=$<1-f_o/f_c>$
with $f_o$ and $f_c$ being the observed flux and the estimated
continuum. Thus $D_A$ is a measure of the mean depression of the
quasar continuum by the ensemble of \lya absorption lines and whatever metal
lines that are present. These $D_A$ values 
are plotted as open circles in Figure 9. The fact that the value
of $1-D_A$ in the bin closest to $z_{em}$ is significantly above all
others is evidence for the proximity effect. We then perform a number of
simulations as described in section 4 and fold in the proximity
effect by modifying the $N$(H I) of each cloud according to
the equation $N$(H I)'=$N$(H I)/(1+$\omega(z)$), where $\omega(z)$
is the ratio of quasar flux at redshift $z$ to the flux of the
UV background (see Lu et al. 1991 for derivations of the relevant
equations). Note that absorption lines associated with
metal systems are assumed to be unaffected.
 We then adjust the mean intensity of the UV
background at the Lyman limit frequency, $J_{\nu}^{LL}$,
until the $D_A$ distributions from the simulated spectra
match that from the observed spectrum. The results are shown in
Figure 9, where the dotted lines indicate the $D_A$ values estimated from 
25 simulations with $J_{\nu}^{LL}=2\times 10^{-22}$ erg s$^{-1}$
cm$^{-2}$ Hz$^{-1}$ sr$^{-1}$. The flux of
Q 0000$-$26 at the Lyman limit frequency is estimated to be
$10^{-27}$ erg s$^{-1}$ cm$^{-1}$ Hz$^{-1}$ from 
the spectrum published by Sargent, Steidel, \& Boksenberg (1989). 
We have adopted an emission redshift of 4.127
for Q 0000$-$26, which is the redshift of the highest-redshift \lya
line detected in the spectrum other than the associated metal system
at $z=4.133$. This redshift agrees well with the value,
$z_{em}=4.124$, estimated from the low ionization emission lines of
O I $\lambda$1302/Si II $\lambda$1304 (Brian Espey, private communication). 
Studies (cf. Gaskell 1982; Espey et al. 1989) have shown that emission 
lines of low ionization species provide a better indication of the 
systemic redshift of quasars than high ionization lines.
The adopted $z_{em}$
is approximately 1000 km s$^{-1}$ higher than the published redshift
of 4.111 by Sargent et al. (1989), which was estimated from the
Ly$\alpha$/N~V emission. We conclude that
$J_{\nu}^{LL}\sim 2\times 10^{-22}$ erg s$^{-1}$ cm$^{-2}$
Hz$^{-1}$ sr$^{-1}$ at $z\sim 4.1$ with an uncertainty
of about a factor of 2 based on experimentations with the simulations.

      We see no obvious difference in the distribution of $b$ values 
for clouds near the \lya emission and for clouds elsewhere.

\bigskip
\centerline{\bf 6 DISCUSSION}
\bigskip
\centerline{\it 6.1 The Importance of Numerical Simulations}
\bigskip

    Throughout the course of this work, we are struck by the amount
of information we have gained from analyses of simulated \lya forest
spectra.  Although fitting the
simulated spectra considerably increase the workload, the effort is nicely
rewarded.  The simulations have helped us to
understand the biases in the derived $N$-$b$ relation, to gauge
the reality of the very narrow and the very broad lines, to make
corrections for the incompleteness in the derived column density distribution,
and to derive $J_{\nu}$ from the proximity effect analysis.
Without such simulations, it would have been very difficult to 
interpret some of the results and to derive unbiased Doppler width
and column density distributions. While such simulations may have limited
use at much lower redshifts where the line density is relatively
low and blending is less severe, they are certainly very valuable at the
high redshift studied here.

\bigskip
\centerline{\it 6.2 Evolution of Cloud Properties}
\bigskip

      The index, $\beta=-1.46\pm0.06$, for the H I column density 
distribution of the clouds in $12.3<{\rm log}N<14.5$ 
found here at $3.4<z<4.0$ is identical to 
the value, $\beta=-1.46\pm0.05$ found by Hu  et al. (1995) at $2.5<z<3.1$
based on a similar analysis. Thus there is no obvious
evidence for an intrinsic evolution 
in the column density distribution of the clouds between these redshifts.

     The fact that the power-law distribution of \lya clouds' column
density extends to $N$(H I) at least as low as $2\times10^{12}$ cm$^{-2}$
has important implications for the study of the He II $\lambda$304
absorption for these clouds, which is important for determining the 
ionization level of the absorbing gas. The He II absorption has been
detected toward three directions (Jakobsen et al. 1994; Tytler et al. 1995;
Davidsen et al. 1996). Because the expected ratio of $N$(He II)/$N$(H I)
is in the range of 10-100, the clouds that dominate the He II absorption
are those with $N$(H I) in the range $10^{11}-10^{13}$ \cm2, which do not
dominate the H I absorption. Thus reliable knowledge of the $N$(H I) 
distribution at the low column density end will help to interpret the
He II absorption results (cf. Songaila, Hu, \& Cowie 1995).

      While the $N$(H I) distribution of the clouds does not show any
obvious evolution between $z\sim2.8$  and $z\sim3.7$, there are 
some differences in the Doppler $b$ distributions of the clouds between 
these two epochs.  In particular, both the mean $b$ value 
($\langle b\rangle=$23 km s$^{-1}$)
and the cutoff $b$ value ($b_{cut}=$15 km s$^{-1}$) found in this study
at $\langle z\rangle=3.7$
are smaller than the corresponding values, 28 and 20 km s$^{-1}$,
found by Hu et al. (1995) for $\langle z\rangle=2.8$.
Although our estimate of $\langle b\rangle=23$ km s$^{-1}$  may be
uncertain by 2-3 km s$^{-1}$, simulation results indicate that
$\langle b\rangle$ for our sample of Ly$\alpha$ clouds is very unlikely
to be as larger as 28 km s$^{-1}$. In particular, we consider
the difference in the values of $b_{cut}$ between the
two studies quite robust based on the discussion in \S5.1.

That \lya clouds appear narrower at higher redshifts has been noted before
by Williger et al. (1994), who studied the \lya forest in the $z_{em}=4.5$
quasar BR 1033$-$0327.
The most straightforward interpretation of 
the cutoff $b$ value is that it reflects   the temperature
of the clouds broadened purely by thermal motions.
If so, then the corresponding temperatures for $b_{cut}=20$ and 
15 km s$^{-1}$ are $2.4\times 10^{4}$ K and $1.4\times 10^{4}$ K, respectively.
However, this interpretation may only be partially correct.  
Recent cosmological
simulations (Cen et al. 1994; Petitjean et al. 1995; Zhang et al. 1995;
Hernquist et al. 1996; Miralda-Escude et al. 1996) all indicate that,
at $z\sim 2-4$, \lya
clouds with $N$(H I)$<10^{16}$ \cm2 appear to be associated with 
low density regions of the universe in filaments, sheets, and voids.
In particular, the relatively high column density \lya clouds are 
associated with
sheets and filaments with typical overdensity 
$\rho/{\langle\rho\rangle}\sim 1-10$ where thermal
broadening generally dominates or is comparable to broadening from bulk
motion, while the very low column density clouds are associated with
density enhancement in voids ($\rho/{\langle\rho\rangle}<1$)
for which the absorption is usually broadened by bulk motions
as the gas expands with the Hubble flow (Miralda-Escude et al. 1996). 
Hence the cutoff $b$ value at the low column density end may be governed
by bulk motions rather than by thermal motions.
In addition, the work of
Miralda-Escude et al. (1996) indicates an increase in the mean temperature
of the absorbing gas (at least for $N$(H I)$>13.5$) with decreasing redshift
over the range $2<z<4$ (see their figure 9), 
in qualitative agreement with observations. 
Quantitative comparisons with published cosmological simulation results
are not terribly meaningful as the Doppler width distributions given by
the various groups are determined in ways different from Voigt profile 
fitting.

     There may also be some differences in the clustering properties
of the clouds between $\langle z\rangle=2.8$ and $\langle z\rangle=3.7$. 
Webb (1987) found marginally significant evidence for clustering 
of \lya clouds (ie, those devoid of obvious
metal lines) on velocity scales of 50-150 km s$^{-1}$. Cristiani et al. (1995)
and Hu et al. (1995) also found evidence for clustering on similar scales,
although Rauch et al.  (1992) did not. In particular, Cristiani et al.
showed that the higher column density \lya clouds show stronger clustering
than the lower column density \lya clouds. The presence of weak 
but significant clustering in the line of sight distributions
of \lya clouds and the recent detections of weak C IV absorption associated
with the relatively high column density \lya clouds 
(Cowie et al. 1995; Tytler et al. 1995;
Sargent et al. 1996) suggest that there may be a continuous distribution
in the physical properties of the quasar absorption clouds from the heavy
metal systems to the traditional \lya forest clouds.
However, we do not find any strong evidence
for clustering in our data for clouds of any column density. 
A similar study of \lya
clouds in the spectrum of a $z_{em}=4.5$ quasar  by Williger et al. (1994) 
did not find evidence for clustering either.
These results may suggest that the clustering of \lya clouds at $z\geq4$ is
lower than at lower redshift. 
This interpretation is consistent with results from 
recent hydrodynamic cosmological simulations (Petitjean et al. 1995;
Zhang et al. 1995; Hernquist et al. 1996; Miralda-Escude et al. 1996),
which suggest that \lya clouds at $2<z<4$
arise from filaments and sheets (and sometimes in voids) of low density
material in between collapsed objects. 
Since smaller structures coalesce and form 
progressively larger structures in these kind of models, it may be expected
that the clustering strength of the \lya clouds should increases with
time (ie, decreasing redshift) as the clouds gradually fall towards collapsed
objects (galaxies, cluster of galaxies) which are obviously clustered.
The models can also explain why higher column density clouds cluster more
strongly.  The recent finding that low redshift \lya clouds 
($z\sim 0.5$) show significant
clustering around metal systems (Bahcall et al. 1996) also supports 
this interpretation.

      Finally, our estimated value of $J_{\nu}$ at $z\sim4.1$ is very similar
to that at $z=4.5$ estimated by Williger et al. (1994). 
Giallongo et al. (1996), using a sample of 10 quasar spectra obtained at a
median resolution of 11 km s$^{-1}$, estimated a value of 
$J_{\nu}=(5\pm 1)\times 10^{-22}$ erg s$^{-1}$ cm$^{-2}$ Hz$^{-1}$ sr$^{-1}$
over the redshift range $2.0<z<4.1$. Our estimate of $J_{\nu}$ at $z\sim4.1$
is somewhat lower than the Giallongo et al. estimate, but nontheless may be
consistent with their value since Giallongo et al. did find some evidence
for a drop of $J_{\nu}$ toward higher redshift.
These values are also 
consistent with those expected if quasars are the main sources
of UV ionizing photons at these redshifts (Haardt \& Madau 1996). 
Thus it appears that quasars
can provide all the UV ionizing photons in the intergalactic space at least
up to the redshift of 4.5 or so, and there is no need to invoke other 
significant sources of UV photons.  This drop in $J_{\nu}$ toward 
higher redshift will certainly affect the ionization conditions 
in \lya clouds and in metal absorption clouds. 

\bigskip
\centerline{\bf 7 SUMMARY }
\bigskip

   We present a high resolution (FWHM=6.6 km s$^{-1}$), high S/N ($\sim 30$)
spectrum of the $z_{em}=4.127$ quasar Q 0000$-$26 obtained with the Keck
telescope. Voigt profiles were fitted to the \lya absorption lines in order
to derive the H I column density and Doppler width distributions of the
clouds. Simulated \lya forest spectra with matching characteristics were
also analyzed in order to understand the effects of line blending/blanketing
and noise in the data. The main results, applicable at a mean redshift
of $\langle z\rangle=3.7$, are summarized below.

1.  The column density distribution, 
after corrections for incompleteness resulting from line blanketing, 
is well described by a power law function
with index $\beta=-1.55\pm0.05$ over the column density range
of $12.6<{\rm log}N({\rm H I})<16.0$. A similar fit over the
column density range $12.3<{\rm log}N({\rm H I})<14.5$ yields
$\beta=1.46\pm0.06$, which hints a possible steepening of the
distribution at log$N>14.5$.

2. The Doppler width distribution is consistent with a Gaussian function
with a mean of 23 km s$^{-1}$ and a dispersion of 8 km s$^{-1}$, but 
truncated at 15 km s$^{-1}$, {\it ie}, no clouds with $b<15$ km s$^{-1}$ are 
required to describe the data.

3. There is no significant  evidence for an intrinsic  correlation between 
the Doppler width and H I column density of the clouds. 
The relative lack of clouds with low-$N$ and high-$b$ in the
observed distribution (figures 3 and 4), which creates
an apparent correlation between $N$ and $b$, can be understood
in terms of measurement biases (\S5.1). On the other hand, if
indeed there is an intrinsic lack of Ly$\alpha$ clouds with 
low-$N$ and high-$b$, such an effect will be difficult to recognize.

4. While the H I column density distribution found here is consistent with
that derived from similar studies at lower redshifts (cf. Hu et al. 1995), 
both the mean Doppler width (23 km s$^{-1}$) and the cutoff value
(15 km s$^{-1}$) are lower than the similar values determined by
Hu et al. for lower redshift clouds (28 km s$^{-1}$ and 20  km s$^{-1}$, 
respectively, at $\langle z\rangle=2.8$). Thus clouds are
on average cooler at the higher redshifts studied here.

5. For clouds with $13.0<$log $N$(H I)$<15.0$,
We find a marginally significant ($\sim 2\sigma$) clustering signal
in the two-point correlation function 
in the velocity interval $100<\Delta v<160$ km s$^{-1}$.

6. Analyses of the proximity effect indicate a value of 
$J_{\nu}\sim2\times10^{-22}$ erg s$^{-1}$ cm$^{-2}$ Hz$^{-1}$
sr$^{-1}$ for the mean intensity of the metagalactic
UV ionizing background at $z\sim 4.1$, which is consistent with
the interpretation that quasars provide the bulk of UV ionizing photons
at this high redshift.

\vskip 1.0 truein
We are indebted to Bob Carswell for kindly providing the VPFIT software
used for this project. We thank Tom Barlow for help
in reducing the data with his own software, Brian Espey for communicating
the emission redshift of Q 0000$-$26, and Michael Rauch for 
useful  discussion and for commenting on an earlier draft of the paper.
We also thank the referee, Greg Bothun, for many helpful comments.
The W. M. Keck Observatory is operated as a scientific partnership 
between the California Institute of Technology and the University of
California; it was made possible by the generous financial support of
the W. M. Keck Foundation. We thank Steven Vogt and the HIRES team for 
building the HIRES spectrograph, and the observatory staff for expert
assistance with the observations.
Support for this work was provided by NASA through grant numbers 
HF1062.01-94A (LL) and HF1040.01-92A (DSW) from the
Space Telescope Science Institute, which is operated by the Association
of Universities for Research in Astronomy, Inc., for NASA under contract
NAS5-26555. WWS acknowledges support from NSF grant AST92-21365.
M. Takada-Hidai appreciates support from the General Research Organization,
Tokai University, which made it possible for him to stay at Caltech during
September 1995 - March 1996.

\eject
\centerline{\bf REFERENCE}
\bigskip

\hangpara
Bahcall, J.N. et al., 1993, ApJS, 87, 1

\hangpara
Bahcall, J.N. et al., 1996, ApJ, 457, 19

\hangpara
Bajtlik, S., Duncan, R.C., \& Ostriker, J.P. 1988, ApJ, 327, 570

\hangpara
Carswell, R.F., in QSO Absorption Lines: Probing the Universe, eds.
    J.C. Blades, D.A. Turnshek, \& C.A. Norman (Cambridge University
    Press), p.91

\hangpara
Carswell, R.F., Webb, J.K., Baldwin, J.A., \& Atwood, B. 1987, ApJ, 319, 709

\hangpara
Cen, R., Maralda-Escude, J., Ostriker, J.P., \& Rauch, M. 1994, ApJ, 437, L9

\hangpara
Cowie, L.L., Songaila, A., Kim, T-S, \& Hu, E.M. 1995, AJ, 109, 1522

\hangpara
Cristiani, S., D'Odorico, S., Fontana, A., Giallongo, E., \& Savaglio,
   S. 1995, MNRAS, 273, 1016

\hangpara
Davidsen, A.F., Kriss, G.A., \& Zheng, W. 1996, Nature, submitted

\hangpara
Espey, B.R., Carswell, R.F., Bailey, J.A., Smith, M.G., \& Ward, M.J. 
  1989, ApJ, 342, 666

\hangpara
Gaskell, C.M. 1982, ApJ, 263, 79

\hangpara
Giallongo, E., Cristiani, S., D'Odorico, S., Fontana, A., \& Savaglio, 
   S. 1996, preprint

\hangpara
Haardt, F., \& Madau, P. 1996, ApJ, in press

\hangpara
Hernquist, L., Katz, N., Weinberg, D.H., \& Miralda-Escude, J. 1996, ApJ, 
in press

\hangpara
Hu, E.M., Kim, T-S, Cowie, L.L., \& Songaila, A. 1995, AJ, 110, 1526

\hangpara
Jakobsen, P., Boksenberg, A., Deharveng, J.M., Greenfield, P.,
   Jedrzejewski, R., \& Paresce, F 1994, Nature, 370, 35

\hangpara
Lu, L. 1991, ApJ, 379, 99

\hangpara
Lu, L., Wolfe, A.M., \& Turnshek, D.A. 1991, ApJ, 367, 19

\hangpara
Miralda-Escude, J., Cen, R., Ostriker, J.P., \& Rauch, M. 1996, ApJ, 
submitted

\hangpara
Petitjean, P., Mucket, J.P., \& Kates, R.E. 1995, A\&A, 295, L9

\hangpara
Petitjean, P., Webb, J.K., Rauch, M., Carswell, R.F., \& Lanzetta, K. 1993
   MNRAS, 262, 499

\hangpara
Rauch, M., Carswell, R.F., Chaffee, F.H., Foltz, C.B., Webb, J.K.,
   Weymann, R.J., Bechtold, J., \& Green, R.F. 1992, ApJ, 390, 387

\hangpara
Rauch, M., Carswell, R.F., Webb, J.K., \& Weymann, R.J. 1993, MNRAS,
   260, 589

\hangpara
Sargent, W.L.W., Boksenberg, A., \& Steidel, C.C. 1988, ApJS, 68, 639

\hangpara
Sargent, W.L.W., Steidel, C.C., \& Boksenberg, A. 1989, ApJS, 69, 703

\hangpara
Sargent, W.L.W., Young, P.J., Boksenberg, A., \& Tytler, D. 1980,
   ApJS, 42, 41

\hangpara
Sargent, W.L.W., Womble, D.S., Barlow, T.A., \& Lyons, R.S. 1996,  
  in preparation

\hangpara
Songaila, A., Hu, E.M., Cowie, L.L. 1995, Nature, 375, 124

\hangpara
Steidel, C.C. 1990, ApJ, 72, 1

\hangpara
Tripp, T.M., Lu, L., \& Savage, B.D. 1996, ApJS, 102, 239

\hangpara
Tytler, D., Fan, X-M, Burles, S., Cottrell, L., Davis, C., Kirkman,
   D., \& Zuo, L. 1995, in  Quasar Absorption Lines, ed. G. Meylan
   (Springer-Verlag), p.289

\hangpara
Vogt, S. 1992, in High Resolution spectroscopy with the VLT,
   ed. M.-H. Ulrich (Garching:ESO), 223

\hangpara
Webb, J.K. 1987, in Observational Cosmology, eds. A. Hewett,
G. Burbidge, \& L.Z.Fang (Reidel, Dordrecht), p803

\hangpara
Williger, G.M., Baldwin, J.A., Carswell, R.F., Cooke, A.J., Hazard, C.,
   Irwin, M.J., McMahon, R.G., \& Storrie-Lombardi, L.J. 1994, ApJ, 428, 574

\hangpara
Zhang, Y., Anninos, P., \& Norman, M.L. 1995, ApJ, 453, L57

\eject
\centerline{\bf FIGURE CAPTIONS}
\bigskip

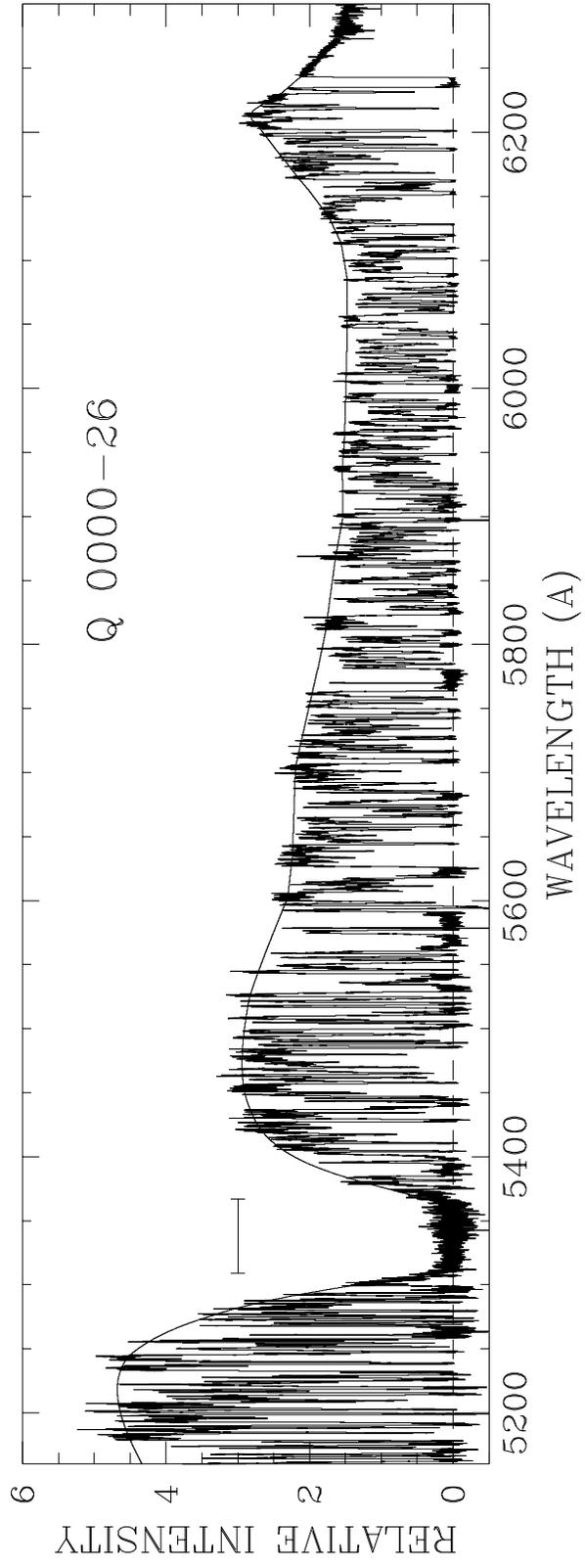
\noindent {\bf Figure 1} Keck HIRES spectrum of Q 0000$-$26 in arbitrary
flux units. Only the \lya forest portion is shown in order to illustrate
the adopted continuum level (smooth solid curve). 
Note the damped \lya absorption at $z_{abs}=3.39$ or $\lambda=5337$ \AA.
The spectral region blocked out by the damped Ly$\alpha$ absorption
(indicated by the horizontal bar) will be omitted from
Figure 2.

\bigskip
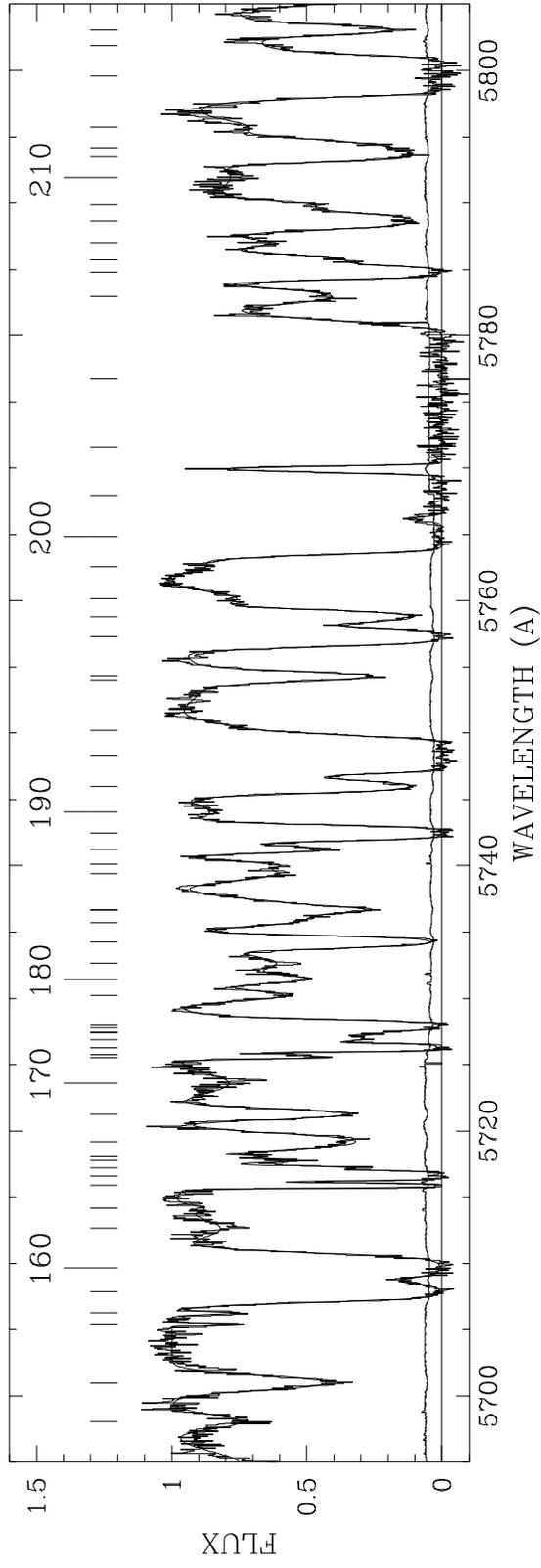
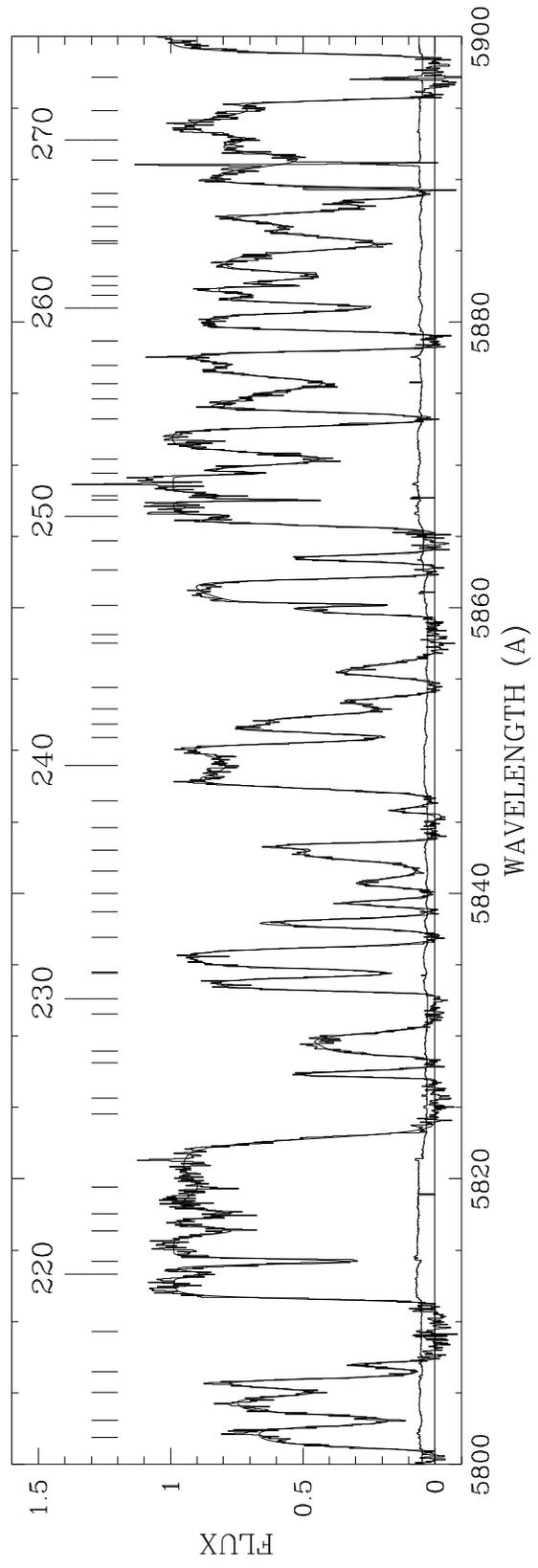
\noindent {\bf Figure 2} Continuum-normalized Keck HIRES spectrum of
Q 0000$-$26. The lower spectrum in each panel is the 1$\sigma$ error
spectrum. Absorption lines at $\lambda>5380$ \AA\ are marked and
listed in Table 1. The results of VPFIT to the Ly$\alpha$ forest
region between 5380-6250 \AA\ are shown as smooth curves superimposed
on the spectrum. Note that the damped \lya absorption at $z_{abs}=3.39$
has been removed. The spectral region between 5309-5367 \AA\ is not 
shown because it is at the bottom of the damped Ly$\alpha$
absorption where the observed flux is zero. 

\bigskip
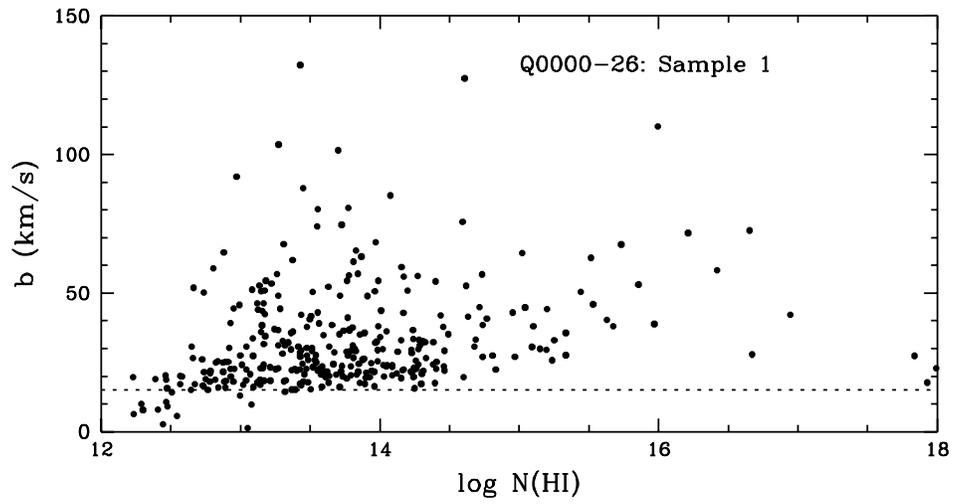
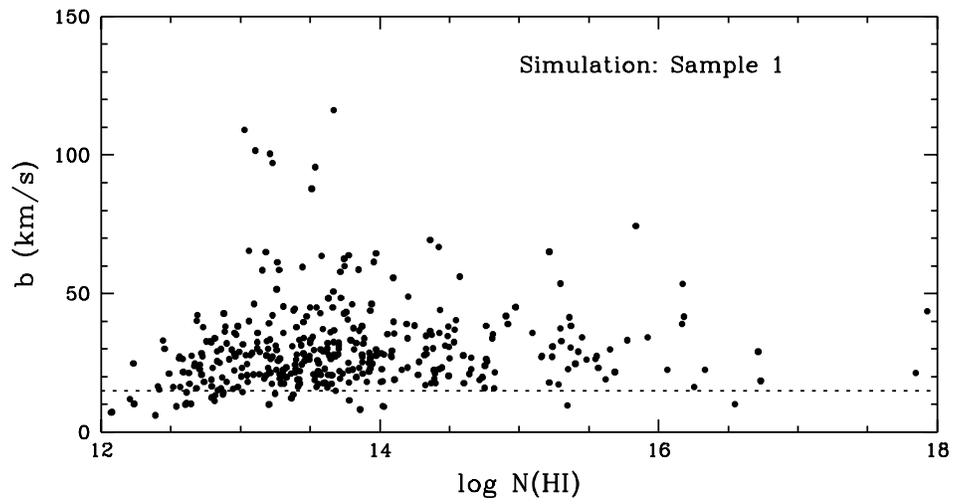
\noindent {\bf Figure 3} Distributions of $b$ vs log$N$ for clouds in
Sample 1 for Q 0000$-$26 (upper panel) and for the simulation (lower
panel). The dashed line represents $b_{cut}=15$ km s$^{-1}$. 

\bigskip
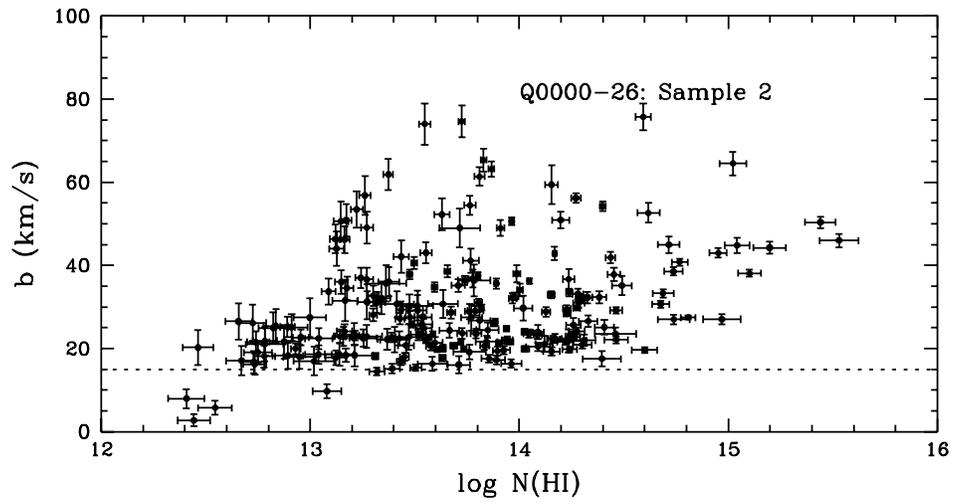
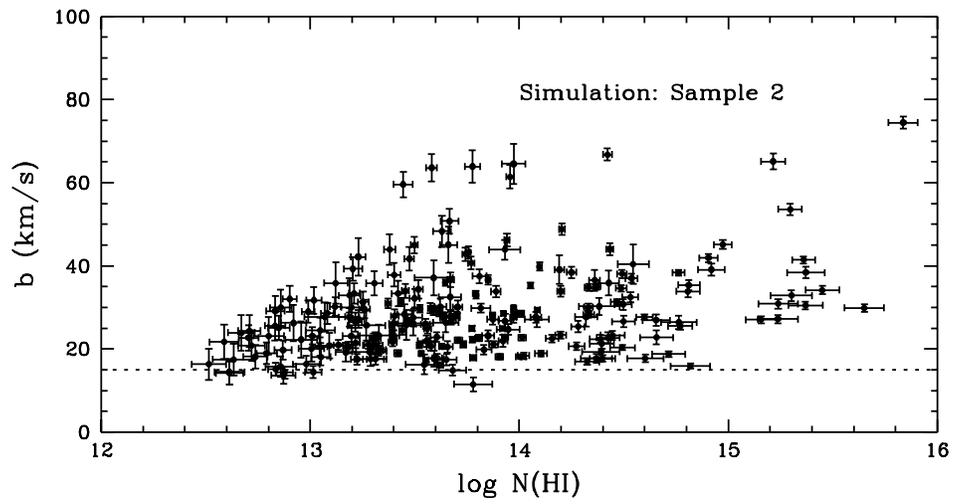
\noindent {\bf Figure 4} Distributions of $b$ vs log$N$ for clouds
in Sample 2 for Q 0000$-$26 (upper panel) and for the simulation (lower
panel). Formal error bars are from VPFIT. The dashed line represents
$b_{cut}=15$ km s$^{-1}$.

\bigskip
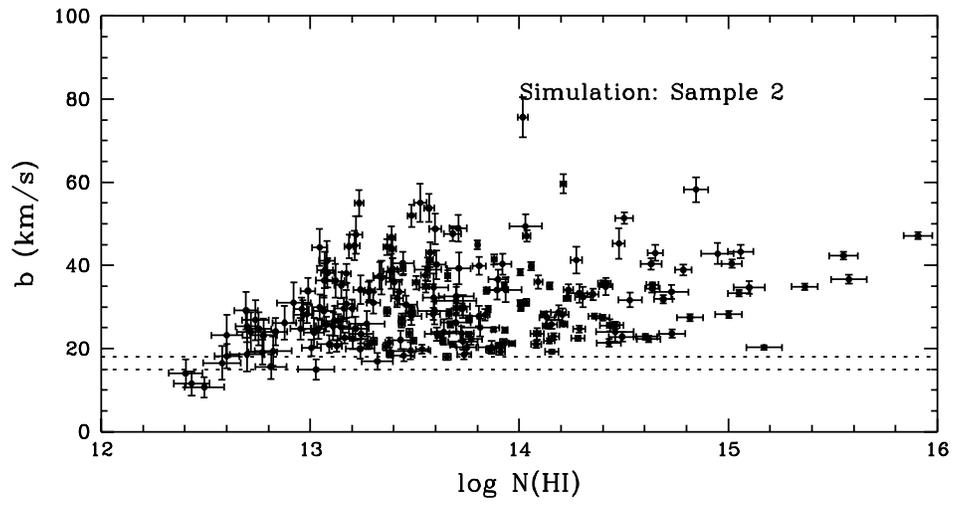
\noindent {\bf Figure 5} Distribution of $b$ vs log$N$ for clouds in
Sample 2 for a simulation where the input $b_{cut}=18$ km s$^{-1}$
rather than 15 km s$^{-1}$. These two cutoff values are indicated by
the dotted lines.

\bigskip
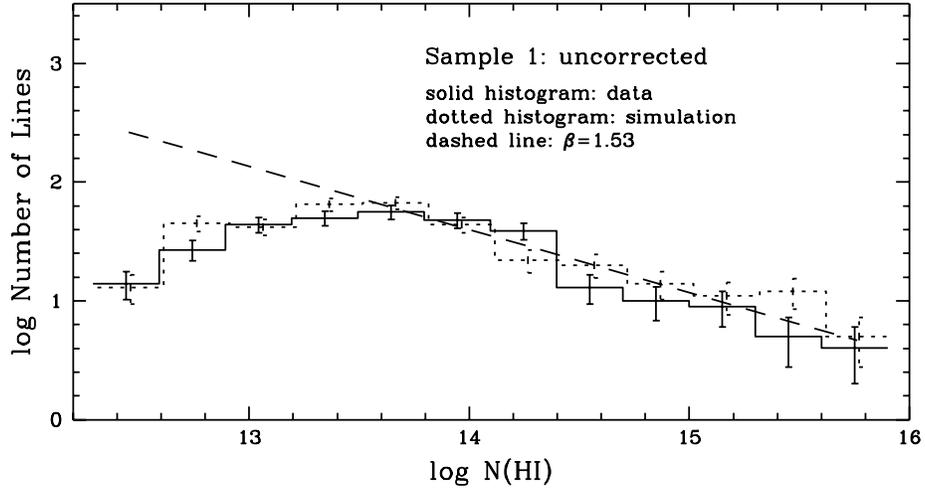
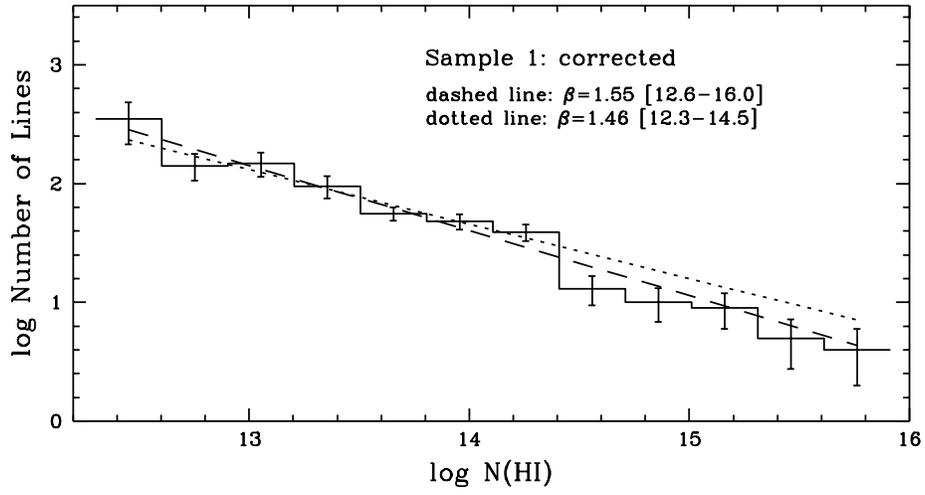
\noindent {\bf Figure 6} Column density distribution of \lya clouds
in Sample 1. The directly measured distributions from profile fitting
are shown as histograms in the upper panel for the observation (solid)
and the simulation (dotted), with the dashed straight line representing
the input distribution to the simulation.  The observed distribution,
after corrections for incompleteness using the simulation results in
the column density range $12.3<$log$N<13.5$, is shown as histogram
in the lower panel, where the dashed line represents the best power-law
fit over the column density range $12.6<$log$N<16.0$ and the
dotted line represents a similar fit over $12.3<$log$N<14.5$.

\bigskip
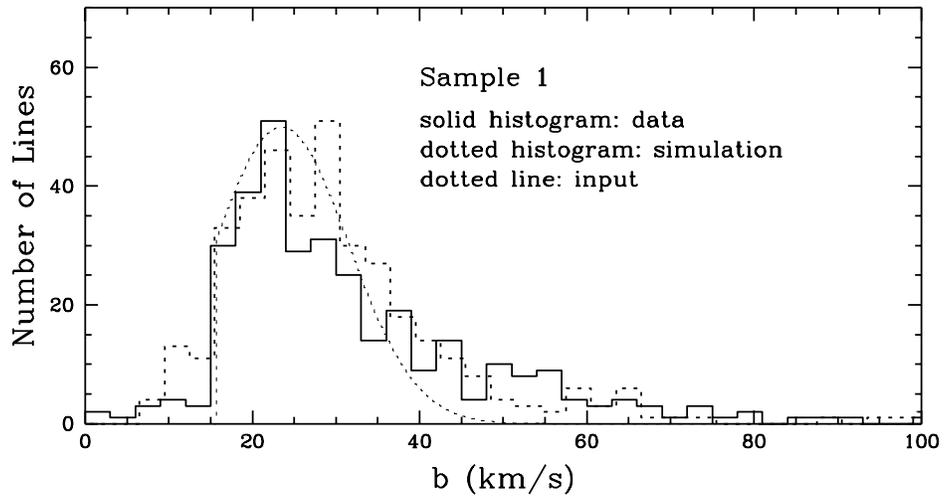
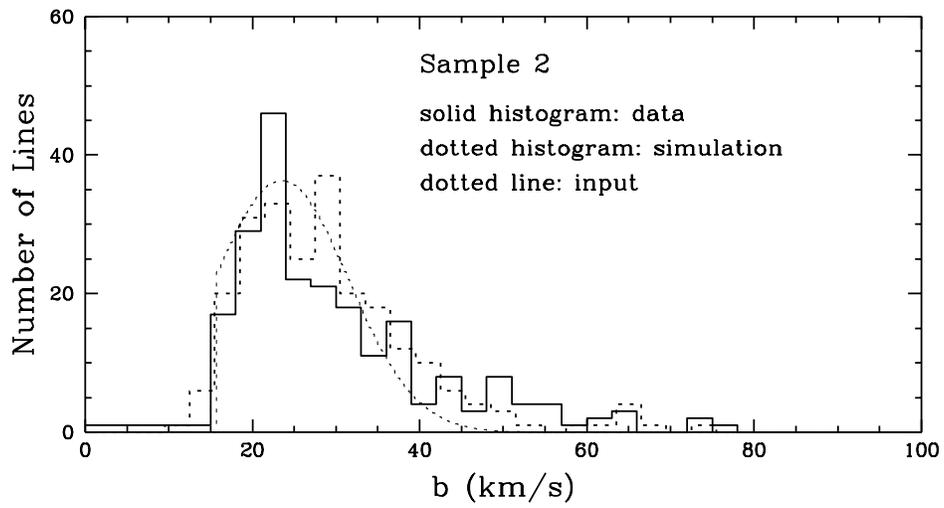
\noindent {\bf Figure 7} Distributions of Doppler width for \lya
clouds in Sample 1 and Sample 2 for the
observation (solid histograms) and for the simulation (dotted histograms).
The dotted smooth curves illustrate the input distribution to the
simulation, which is a Gaussian with a mean of 23 km s$^{-1}$ and 
dispersion of 8 km s$^{-1}$, but truncated at 15 km s$^{-1}$. 
The input distribution curves have been scaled to roughly match
the distributions from profile fitting.
The dotted curve/histogram are shifted slightly in the x-direction
with respect to the solid histogram for clarity.

\bigskip
\noindent {\bf Figure 8} Two-point correlation function for \lya
clouds in Sample 1 for the real data (top panel) and for the
simulation (lower panel). The waving curves indicate the $\pm\sigma$
standard deviation for randomly distributed clouds. 

\bigskip
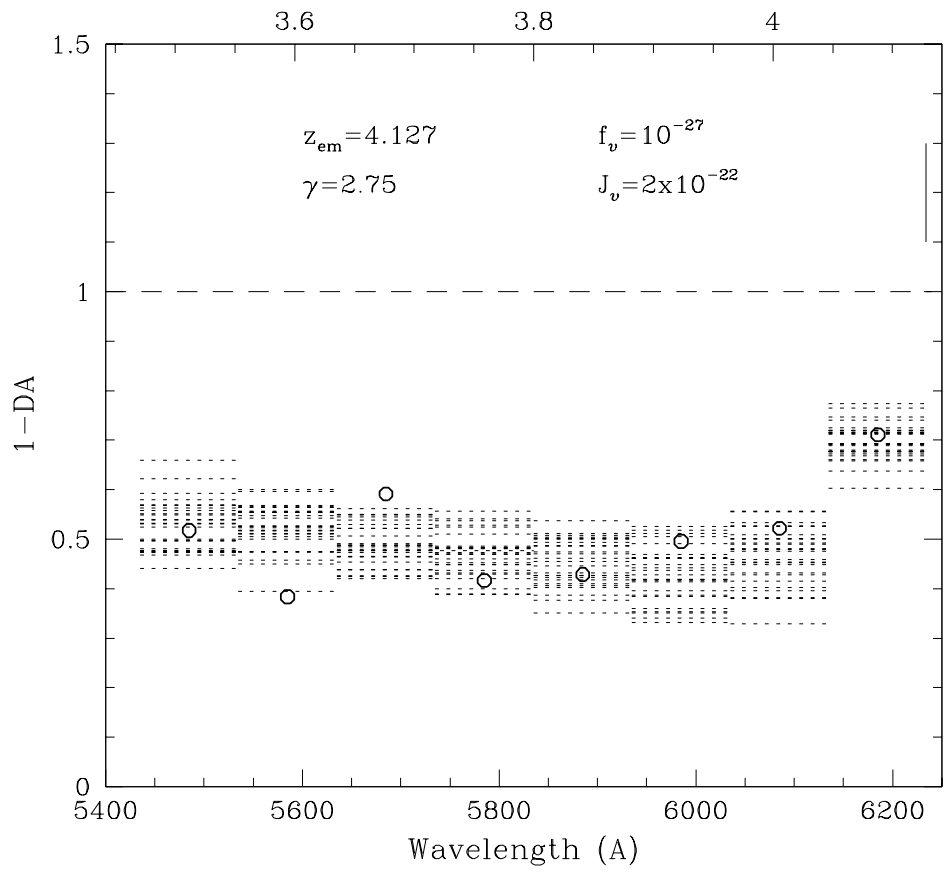
\noindent {\bf Figure 9} Proximity effect analysis. The open circles
represent $D_A$ measurements in each 100 \AA\ bin from the observed
spectrum of Q 0000$-$26, while the dashed lines represent similar
measurements from 25 simulations. The vertical line marks
the adopted redshift for Q 0000$-$26: $z_{em}=4.127$.

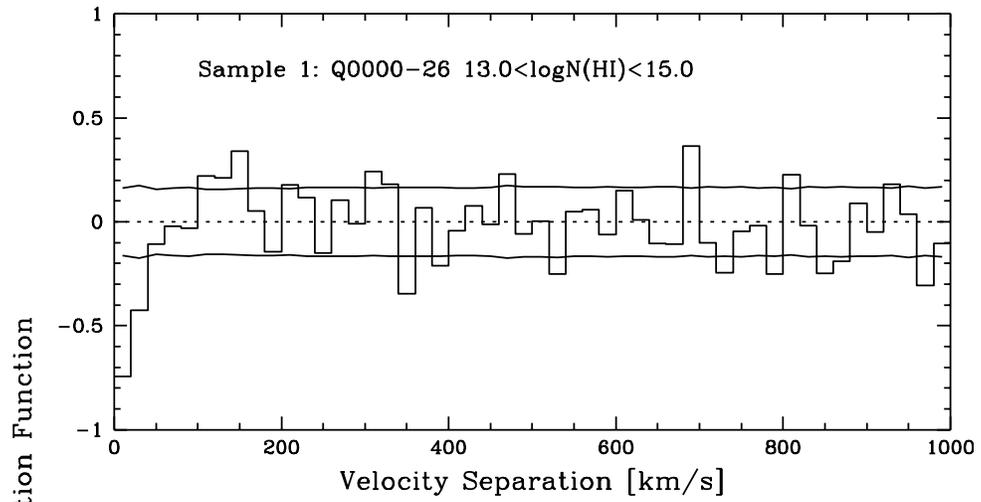
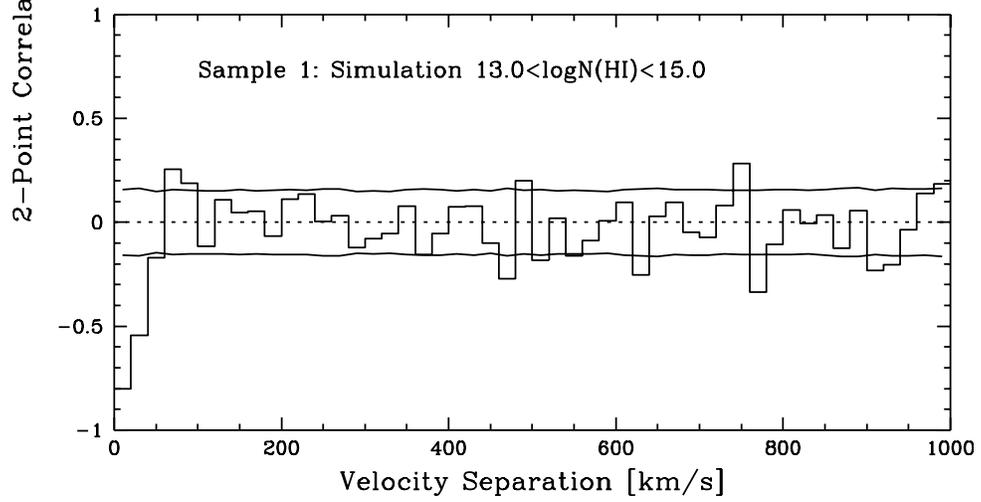
\bye